\documentclass{article}

\usepackage{arxiv}

\usepackage[utf8]{inputenc} 
\usepackage[T1]{fontenc}    
\usepackage{hyperref}       
\usepackage{url}            
\usepackage{booktabs}       
\usepackage{amsfonts}       
\usepackage{nicefrac}       
\usepackage{microtype}      
\usepackage{lipsum}
\usepackage{graphicx}
\usepackage{natbib}
\usepackage{enumitem}

\graphicspath{ {./images/} }

\newcommand{\checkbox}{$\square$}

\title{When Discourse Stalls: Moving Past Five Semantic Stopsigns about Generative AI in Design Research}

\author{
Willem van der Maden \\
HCI \& Design Section\\
  ITU Copenhagen\\
  Copenhagen, Denmark \\
  \texttt{wiva@itu.dk} 
  \And
Vera van der Burg \\
  Designing Intelligence Lab\\
  Technical University Delft\\
  Delft, Netherlands 
  \And
Brett A. Halperin \\
  Human Centered Design \& Engineering\\
  University of Washington\\
  Seattle, Washington, United States 
  \And
Petra Jääskeläinen \\
  EECS/HCT/MID\\
  KTH Royal Institute of Technology\\
  Stockholm, Sweden 
  \And
Joseph Lindley \\
  Imagination\\
  Lancaster University\\
  Lancaster, United Kingdom 
  \And
Derek Lomas \\
  Industrial Design Engineering\\
  TU Delft\\
  Delft, Netherlands 
  \And
Timothy Merritt \\
  Dept. of Computer Science\\
  Human-Centered Computing\\
  Aalborg University\\
  Aalborg, Denmark
}

\begin{document}
\maketitle
\begin{abstract}
This essay examines how Generative AI (GenAI) is rapidly transforming design practices and how discourse often falls into over-simplified narratives that impede meaningful research and practical progress. We identify and deconstruct five prevalent ``semantic stopsigns''---reductive framings about GenAI in design that halt deeper inquiry and limit productive engagement. Reflecting upon two expert workshops at ACM conferences and semi-structured interviews with design practitioners, we analyze how these stopsigns manifest in research and practice. Our analysis develops mid-level knowledge that bridges theoretical discourse and practical implementation, helping designers and researchers interrogate common assumptions about GenAI in their own contexts. By recasting these stopsigns into more nuanced frameworks, we provide the design research community with practical approaches for thinking about and working with these emerging technologies.
\end{abstract}


\section{Introduction}
It has been roughly three years since the open-source release of \textit{Stable Diffusion} ignited a Generative AI (GenAI) boom~\citep{bengesi2023advancementsgenerativeaicomprehensive}. The proliferation of these technologies has since reshaped design practice and research.  From early ideation to final implementation, these developments have significantly altered how design work is conceived, conducted, and evaluated~\citep{hou2024generativeai}. This essay examines the critical juncture at which the design research community finds itself, seeking to understand and shape these developments while grappling with their implications for creative practice, design education, and professional identities. Popular discourse around GenAI often centers on simplified unequivocal narratives: AI as a threat to humanity, as a solution to global challenges, as a force of disruption, or as a replacement for humans~\citep{gilardi2024narratives}.

While these narratives have sparked debate and interest, they can function as ``semantic stopsigns”---conceptual framings that oversimplify complex issues, providing an illusion of resolution that hinders deeper inquiry~\citep{lesswrong, lifton1961thought}. For instance, claims like “AI is unreliable” can lead to outright dismissal of its potential, even in cases where it could support exploratory or creative workflows with proper oversight. Similarly, concerns such as “all GenAI infringes intellectual property” can lead to a dismissal that prevents stakeholders from exploring nuanced strategies for ethical and responsible implementation. Existing literature has provided valuable insights into GenAI’s technical capabilities and ethical concerns, laying a crucial foundation for understanding its potential. However, much of this work remains focused on theoretical questions or specific technical challenges, leaving practice-oriented implications---such as its influence on design education, workflows, and organizational strategies---less explored. For example, reductive assumptions about AI can lead to curricula that either prioritizes technical skills over critical thinking or avoid AI entirely due to excessive fears~\citep{walter2024ai,jaramillo2024ai}. Similarly, organizations often adopt AI tools hastily without evaluation~\citep{mit2023ai} or dismiss them outright~\citep{forbes2024ai}, missing opportunities for meaningful innovation. Addressing these barriers requires moving beyond semantic stopsigns to develop nuanced, practice-oriented knowledge that bridges theoretical discourse with grounded insights~\citep{computationalcreativity-kaila}.

In our reflection on GenAI in design, we embrace critical optimism as our guiding perspective---an approach to technological engagement first explored by Anna~\citet{Brynskov2020} and further developed in social research~\citep{amaturo2021critical}. Rather than adopting either uncritical acceptance or wholesale rejection, critical optimism encourages thoughtful engagement with new technologies through direct experience. This mindset, or "posture" as Amaturo and Aragona describe it, proposes that meaningful understanding of technologies like GenAI emerges through hands-on interaction---experimenting, playing, prototyping, and embedding these tools within practice. While critical optimism shapes our overall approach, we structure our analysis through Theory U~\citep{scharmer2009theory}, identifying five prevalent semantic stopsigns in GenAI discourse and demonstrating how to move beyond them through reflection and reframing.

This paper examines how simplified narratives, or ``semantic stopsigns,'' impede meaningful engagement with GenI in design research. Drawing on two expert workshops at \textit{ACM Designing Interactive Systems Conference} and semi-structured interviews with design practitioners, we identify and analyze five common stopsigns that halt deeper inquiry into GenAI’s role in design. Our analysis contributes to what~\citet{hook2012} describe as \textbf{intermediate-level} knowledge: knowledge forms that are more abstract than specific instances but not as generalized as overarching theories. We deliberately adopt this approach to provide \textit{generative insights}---knowledge that designers and researchers can adapt and apply within their own contexts to inform debates, guide practice, and advance their understanding of GenAI. By revealing the systemic structures behind these stopsigns and highlighting additional dimensions beyond their simplified framings, we aim to enable more nuanced conversations about GenAI’s implications for design research and practice.

\section{Related work}
To situate our research, we explore the technical foundations of GenAI systems, their integration into design practices, and the ethical and practical challenges accompanying their adoption. While these systems have shown immense potential, current discourse often becomes ensnared in simplified narratives that impede deeper engagement and actionable progress. Despite the proliferation of frameworks, principles, and guidelines (\citet{algorithmwatch2020inventory} catalogued 167 examples in 2020), high-level initiatives frequently fall short in practice. For example, \citet{correa2023worldwide} highlight that while such frameworks are ostensibly generalizable, they often rely on unrepresentative samples and face substantial challenges when translated across diverse contexts. In practice, these frameworks tend to produce a “procedural orientation,” where practitioners follow predefined processes rather than cultivating the critical competencies necessary for meaningful engagement with these technologies~\citep{madaio2024learning}.

\subsection{GenAI in design}
Over the past decade, GenAI has evolved from early proof-of-concept algorithms to powerful, commercially viable systems that are reshaping design practices~\citep{unveiling2024}. Initial breakthroughs with Generative Adversarial Networks (GANs) demonstrated the feasibility of producing novel images from latent representations, sparking rapid innovation in creative and design fields~\citep{hughes2021generative}. More recently, diffusion models and large language models have further expanded these capabilities to support multimodal tasks, enabling designers to integrate text, image, and even audio data in unprecedented ways~\citep{zhang2023multimodal}. The practical applications of these technologies span a wide range of design processes~\citep{he2023exploring}, from AI-assisted architectural prototyping~\citep{ko2023experiments} to automated user interface generation~\citep{wu2024uicoder} and algorithmic fashion design~\citep{choi2023developing}. Scholars and industry practitioners particularly emphasize GenAI's capacity for \emph{co-creation}---where designers and algorithms collaboratively generate novel solutions that neither party would have arrived at independently~\citep{zhou2024understanding, 10494260}. The promise of ``faster ideation," wherein AI reduces time-intensive tasks like creating mood boards or generating style variations, suggests designers might be freed to focus on higher-level concerns~\citep{bursztyn2021gaudi,choi2023creativeconnect, choudhury2025promise}. 

However, this transformative narrative faces significant critique and challenges. Some scholars question whether GenAI truly ``transforms" design methods or simply replicates known patterns and aesthetics derived from training data~\citep{mccormack2024prompt}. By learning from existing work, AI systems may inadvertently reinforce dominant cultural trends or stylistic norms, limiting the originality that design practice strives for~\citep{algorithmwatch2023cultural,  wadinambiarachchi2024effects, prabhakaran2022cultural, cetinic2022myth}. There are also concerns that ``surprising" AI outputs may be mistaken for genuine creative innovation, while in reality they result from underlying biases embedded in the datasets~\citep{arxiv2024patterns}.

The field simultaneously grapples with substantial ethical, legal, and environmental challenges. A major line of inquiry examines how AI may perpetuate biases---both social and aesthetic---by reproducing patterns found in the training data~\citep{buolamwini2023unmasking, crawford2021atlas, bender2021dangers}. Intellectual property rights present another critical concern, particularly when AI systems use copyrighted material to generate content that may infringe upon original creators' works~\citep{ducru2024ai, schultz2024generative}. These issues spark ongoing debates about what constitutes ``fair use" of training datasets and how designers or organizations should navigate the risk of IP infringement~\citep{dzuong2024uncertain, murray2023generative}. Similarly, environmental impact has emerged as an additional crucial consideration, with researchers highlighting the significant computational overhead of training and running large-scale models~\citep{strubell2019energy, liu2024green}. Discussions about energy usage, e-waste, and resource-intensive training pipelines underscore the need for more sustainable AI techniques~\citep{le2024recommendations, ALZOUBI2024143090}. These varied ethical, legal, and environmental concerns appear as recurring themes in GenAI literature, forming the core of what some label ``big narratives" in the field~\citep{gilardi2024narratives}.

\subsection{Limited progress in addressing repetitive narratives}
Despite the recurring acknowledgment of these challenges, the discussion around GenAI in design has tended to remain cyclical. Scholars frequently revisit the same themes---such as potential job displacement, biased outputs, or IP ambiguities---yet struggle to push beyond familiar talking points~\citep{gilardi2024narratives}. This pattern reflects a broader challenge in AI literacy discourse, where accessible tools based on real examples are increasingly proposed as solutions to the limitations of fragmented, high-level frameworks and one-size-fits-all approaches~\citep{pinski2024ai}. Yet developing effective AI literacy requires an interplay between practical knowledge, professional judgment, and theory---a gap that current high-level frameworks struggle to address~\citep{sperling2024search}. This repetitive discourse can inadvertently stifle innovation, as researchers and practitioners find themselves rehashing similar debates---about bias, IP, and environmental impact---without a clear path toward deeper inquiry. Recent work suggests that Research-through-Design (RtD) may offer a way forward, functioning as a ``rapid response methodology" uniquely positioned to engage with emerging AI technologies while their socio-cultural ramifications are still being negotiated~\citep{benjamin2024responding}. Such practice-based approaches could provide the missing bridge between abstract understanding and real-world application that the field currently lacks.

\section{Method}
This paper follows a format of a critical essay~\citep{bardzell2010hci, bardzell2016humanistic} and has been informed by various discourse with design research practitioners. Through this approach, the essay serves as a tool for constructing knowledge, provoking engagement, and enabling critical interpretation. Rather than reiterating well-worn summaries of GenAI's capabilities and limitations, we explore why certain narratives about its use in design remain entrenched and how the field might navigate new possibilities. Our goal is to offer ``mid-level knowledge"~\citep{hook2012}---practical insights that bridge the gap between abstract theory and technical implementation. 

To produce this critical essay, we drew inspiration from the Delphi method's approach to iterative expert consultation, which has previously shown merit in HCI research~\citep{jones-2015, alexander2018grand}. Over a 22-month period (October 2022-August 2024), we gathered insights from the GenAI design research community through two expert workshops at ACM conferences (16 and 23 participants)~\citep{van2023towards,van2024death} and semi-structured interviews with five senior design practitioners whose backgrounds span architectural design, educational games, service design, and parametric modeling. Workshop participants brought diverse perspectives from HCI, design, AI, theater, artistic practice, and software development. These empirical insights, combined with our own experiences working with GenAI in design research projects, provided a rich foundation for examining how semantic stopsigns manifest across different design contexts. Rather than presenting these sources separately, we synthesize them into five key narratives, drawing inspiration from Theory U~\citep{scharmer2009theory} as our analytical framework. 

\begin{figure}[h!]
    \centering
    \includegraphics[width=0.8\linewidth]{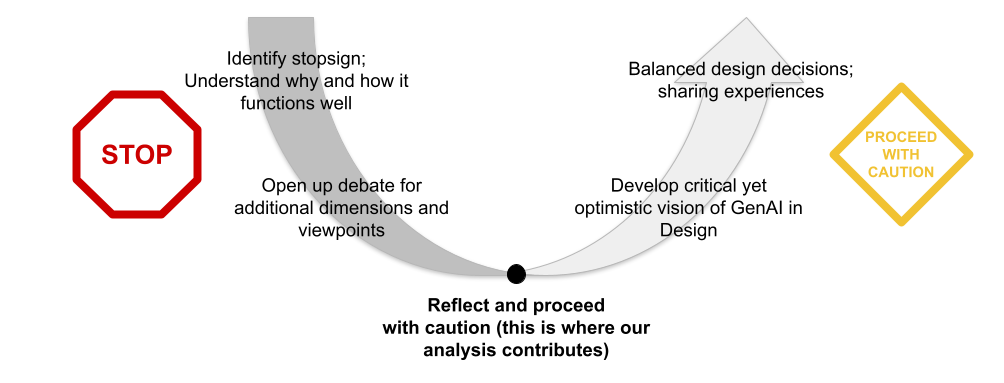}
    \caption{This figure maps the process of moving beyond semantic stopsigns in GenAI discourse, following Theory U’s stages~\citep{scharmer2009theory}: identify stopsigns (downloading), explore new dimensions (sensing), and reflect to crystallize actionable insights (presencing). This process encourages balanced design decisions and a critical yet optimistic vision of GenAI in design. Whether you ultimately embrace or reject GenAI, this approach helps ground your decision in thoughtful reflection rather than fear or oversimplified narratives.}
    \label{fig:u}
\end{figure}

Theory U is a framework typically used in change management contexts to help individuals and organizations navigate complex challenges and transitions. It describes a process of transformation, beginning with recognizing and suspending past patterns (“downloading”), to enable deeper sensing and fresh perspectives, ultimately leading to generative action and innovation. Similarly, our analysis follows a four-step process: (1) identifying the semantic stopsign, (2) understanding how it functions to halt further inquiry, (3) opening up alternative dimensions and perspectives to encourage deeper engagement, and (4) reflecting on these insights to crystallize actionable ideas that can be applied to one's own (design) context. At this stage, one is equipped to begin prototyping new approaches and solutions, see Figure \ref{fig:u}.

Each narrative examines a common \textit{semantic stopsign}—a concept or framing that tends to halt further inquiry by providing an illusion of resolution without deeper exploration. These stopsigns often emerge as natural responses to complex technological change, echoing the “downloading” of past patterns described by \citet{scharmer2009theory}. For each stopsign, we explore:
\begin{enumerate}
    \item A real-world or hypothetical scenario showing how the stopsign manifests in practice
    \item An examination of the stopsign's function in discourse, including how it may truncate deeper discussion
    \item An exploration of the additional dimensions and considerations that lie beyond the initial framing
\end{enumerate}

Importantly, our analysis does not aim to be comprehensive or definitive. Rather, it represents an initial effort to identify and unpack some of the common blockages in current GenAI discourse, with the goal of enabling richer theoretical, practical, and public conversations about these technologies. By providing this scaffolding for deeper engagement, we hope to contribute to more thoughtful and productive ways of understanding and working with GenAI in design practice.

\section{Semantic stopsigns}
This section explores five common stopsigns in GenAI discourse: perceptions of unreliability, intellectual property concerns, the framing of AI as a mere tool, environmental criticisms, and fears of economic degradation. By examining these narratives, we aim to foster deeper inquiry and more constructive engagement.

\subsection{``AI is unreliable''}\label{sec:unrel}
A recent example comes from an architectural firm experimenting with AI-generated concepts for ideation of complex structures such as facades and stairwell designs~\citep{li-2024,barker-2023}. Although the AI-produced renderings were visually striking—evoking the fluid, futuristic style of Zaha Hadid’s work—many of the depicted structures were impossible to build. The AI system had generated curved landings that defied engineering constraints and balustrades attached to non-existent floors, ultimately showcasing visually stunning but unbuildable illusions. Similar issues arise in other creative domains where AI-generated imagery can overlook important details: think of AI-generated concept art for user interfaces with missing or duplicated elements, or photographs of people with extra fingers. Despite these flaws, the firm’s senior architects expressed an immediate reservation toward the AI designs, dismissing them outright as “unreliable.” Once this label was attached, the conversation quickly moved on, halting any deeper evaluation of how such AI tools might be refined and integrated.

A statement like “AI is unreliable” can operate as a semantic stopsign precisely because it leverages widespread awareness of AI’s current limitations---such as hallucinations, biases, and the generation of non-functional designs---and uses them as grounds to dismiss the entire technology without further inquiry. In professional design research and practice, where standards and customs have been honed over decades, there is understandable concern about introducing tools that could compromise quality or safety. Many design professionals have a vested interest in upholding the rigor and reliability of existing processes, which leaves them reluctant to integrate technologies that threaten to disrupt established workflows or require new ways of working. Once the accusation of “unreliable” is voiced, it often carries an air of finality, preventing conversation about context-specific strategies, balancing human oversight with AI-based ideation, or exploring incremental pathways to test and improve AI capabilities within well-defined project constraints.

While the concerns behind calling AI “unreliable” are valid, using that label as a wholesale dismissal overlooks the complexity of AI’s role in design. Yes, AI tools can produce baffling errors and biases---but with proper understanding and structured oversight, designers can harness these systems for tasks ranging from exploratory ideation to detailed design support. In many instances, AI might serve as a starting point, offering a range of generative possibilities that human designers can then refine and adapt to real-world constraints. Taken as a whole, the outcome of a generative image model may be unsuitable for construction; nevertheless, parts or elements of generated imagery can still provide useful inspiration.

That said, it remains critical to exercise caution and maintain a neutral, informed stance: AI is neither a magic bullet nor a hopeless liability. By acknowledging both the pitfalls (e.g., ``hallucinations,” bias, flawed outputs) and the potential (e.g., rapid ideation, pattern recognition, computational power), design researchers can chart middle-ground approaches. This might include developing ethical guidelines for AI in design, conducting rigorous user testing, refining prompt-engineering techniques, and establishing robust validation processes to catch errors early. When handled thoughtfully, AI becomes another powerful tool in the designer’s toolkit---one that, like any new technology, must be understood, shaped, and continually evaluated to ensure it meets the high standards of professional design research and practice. This brings up recurrent themes about how designers, and architects are increasingly responsible for larger portions of the design process. Kolarevic explained this as the ``master builder,''~\citep{kolarevic2003architecture} which, as CAD and digital tools emerged and new blobby or complex designs could be drawn, the architect had to get dirty and engineer the production process along with construction companies. GenAI’s unreliability is a challenge, not a verdict. By refining workflows, building skills, and maintaining critical oversight, designers can turn flaws into opportunities for innovation.

\subsection{``All GenAI infringes IP''}
Imagine a multi-disciplinary team working on a project involving, for example, design, software engineering, physical prototyping, UX research. We can envisage different types of GenAI that could be used at different points in the project. For example, LLMs could be used for analysing text to expedite background research and writing code to assist the software engineers. Meanwhile, image diffusion models could be utilised by designers for creating early-stage visualisations of user interfaces and physical designs or to produce materials for use in UX research. Of course, modern GenAI models are frequently multi-modal, agentic frameworks are used, and multiple models can be used concurrently, but these examples are intended to show there could be relevance across a multidisciplinary team. Let’s imagine the team is discussing with the client whether or not to use GenAI in these ways, in this context anyone in the discussion (i.e., on the team, or the client, or some other observer) could invoke the notion ``Any use of Generative AI is plagiarism" and it could function as a difficult-to-challenge semantic stopsign.

This stopsign resonates because we find ourselves in a situation where training data is inseparable from model capabilities, yet no consent was ever gained for that type of use. For example, if an artist shared digital representations of their work on a blog site that later became part of a model's training set, elements of their work become inseparable from the model's capabilities without explicit consent. It is sometimes possible for data included in training sets to be recreated almost pixel-for-pixel or word-for-word; more commonly, generative AI produces outputs that are similar in style to original works. When these models can generate qualitatively comparable outputs, it raises valid concerns about undermining creative professions from scriptwriters to cartoonists, especially given intellectual property law that predates such technology.

These complexities mirror long-standing discussions in design about inspiration, iteration, and attribution. Proponents argue that training on publicly available data constitutes fair use, particularly since generated outputs are transformative rather than direct copies~\citep{charlesworth2024generative}. Critics counter that while the data isn't copied like-for-like, the resulting models produce outputs which achieve the same effect as what a copy would have done, raising fundamental questions about authorship and attribution~\citep{alkfairy2024ethical}. This tension extends beyond AI to broader debates in digital culture, from 1980s remix culture through to a cornucopia of infamous legal battles over creative ownership~\citep{lessig2008remix}.

Rather than letting these concerns halt innovation entirely, design teams can adopt pragmatic approaches to working with AI while respecting creative rights. This might include developing clear attribution practices, documenting AI usage transparently, and engaging proactively with creators whose work may be referenced. Organizations like the Free Software Foundation suggest evolving IP frameworks to balance creator rights with broader goals of innovation, collaboration, and equitable access~\citep{FSF_copyright_2021}. By acknowledging these complexities while seeking constructive solutions, teams can harness AI's potential while contributing to the development of more nuanced frameworks for creative practice in the digital age.

\subsection{``AI is just a tool''} 
The framing of AI as “just a tool” persists in design discourse, offering reassurance amid anxieties about automation and creative autonomy. Consider a design studio adopting GenAI for rapid ideation: the director dismisses concerns with, “It’s just another tool, like switching from Photoshop to Figma.” Months later, the team’s concepts increasingly mirror algorithmic trends---their once-distinctive style diluted by Midjourney’s bias toward polished, hyper-detailed aesthetics. Conversely, a strategic consultancy resists AI integration, with a senior designer arguing, “It’s just hype; it can’t replace real design thinking.” When clients demand AI-driven user research synthesis, the team struggles to adapt, losing ground to competitors.

This metaphor resonates because GenAI systems are technically bounded, generating outputs within the confines of their training data~\citep{Gillespie-2024}. Like any technology, they serve tasks within workflows----a kernel of truth that reassures practitioners facing AI’s perceived threats to expertise. For adopters, framing AI as a tool provides psychological safety, positioning it as familiar and controllable. For skeptics, dismissing it as “just hype” justifies disengagement, shielding them from the discomfort of evolving skillsets or ethical dilemmas. Yet this reassurance comes at a cost: it obscures how AI’s technical constraints actively mediate outcomes. 

For adopters, the “just a tool” narrative masks how AI’s generative logics—training data biases, output constraints—steer outcomes. \citet{latour2005reassembling} described this as actancy, the capacity of technologies (and tools) to mediate outcomes through their material properties and affordances affordances. While all tools exhibit actancy (e.g., Photoshop’s pen tool nudges designers toward vector precision), GenAI amplifies it uniquely. Unlike tools like Photoshop or CAD, which rely on human-provided assets (e.g., brushes, geometries), GenAI generates material wholesale from latent data patterns. Photoshop and CAD constrain \textit{how} designers work; GenAI shapes \textit{what} they work with, embedding training corpus assumptions into concepts and aesthetics. A designer struggling with Photoshop’s pen tool retains agency over execution; one using DALL-E inherits its parametric bias, steering projects toward computational styles~\citep{ko2023experiments}.

For skeptics, dismissing AI as “just hype” overlooks its potential to shape not only design practice but creatviity as a wohle. While current systems are narrow, their ability to augment ideation, automate repetitive tasks, and synthesize complex data is undeniable. A service design team avoiding AI-driven research tools might miss efficiency gains while failing to interrogate how those tools could amplify biases in their datasets. This avoidance risks leaving teams unprepared for AI’s growing role in design ecosystems.

Moving past this stopsign requires acknowledging AI’s relational influence---its capacity to reshape practice even within technical bounds. Teams might audit how tools like Stable Diffusion privilege certain aesthetics~\citep{cetinic2022myth}, fine-tune models on bespoke datasets to align with their studio’s ethos~\citep{zhou2024understanding}, or explore the effets of homogenization can be used to overcome design fixation~\citep{hoggenmueller2023creative}. For example, a sustainability-focused firm could train a diffusion model on eco-friendly architectures, intentionally shaping its ``creative DNA'' to reinforce their identity. Such approaches treat AI not as a static tool but as a collaborative medium---one that reflects and refines a team’s values through iterative engagement.

\subsection{``AI is bad for the environment''} \label{sustainability}
Consider a research group developing AI models to monitor endangered species populations. Their system processes vast amounts of sensor data to track animal movements and predict population changes, requiring significant computational resources. When questioned about the environmental cost of running their models, a senior researcher dismisses concerns by stating that ``AI's carbon footprint doesn't matter compared to the species extinctions we're preventing". Meanwhile, at a design conference workshop, participants reject proposals to use GenAI tools in sustainable design projects, arguing that any AI use inherently contradicts environmental goals. These two cases exemplify how environmental concerns can function as semantic stopsign both ways, exemplifying the paradoxes that can arise in discourse around environmental sustainability of GenAI and preclude deeper examination of when and how GenAI might be thoughtfully applied.

This semantic stopsign resonates because it builds on legitimate concerns about technology's environmental impact. However, problems arise when these legitimate sustainability concerns -- for example raised in HCI~\citep{nardi2018, bornes2024} and beyond~\citep{Raworth2017-lw} -- are used as discourse startegies to close down the conversation about actual solutions and ways forward. AI systems' resource requirements—from data center energy consumption to hardware production—present real environmental costs that can't be dismissed~\citep{jaaskelainen_environmental_2022, jaask_nordichi2022}. These costs can be hard to determine, as for instance, AI might increase short-term energy use while enabling longer-term sustainability gains through optimized resource management or environmental monitoring, or vice versa. This reveals an important paradox: AI systems developed for environmental conservation might themselves contribute to environmental harm through their computational demands and species surveillance implications. Rather than using this paradox to halt discussion, it can serve as a prompt for more nuanced evaluation of when AI's environmental costs are justified by meaningful conservation outcomes.

Furthermore, the compexity of the environmental concerns extends beyond simple carbon footprint calculations to AI's role in more-than-human systems~\citep{jaaskbook-xd, computationalcreativity-jaask, chiletbook-xd, Nicenboim_Giaccardi_Redström_2023}. 
Moving beyond the stopsign requires examining specific contexts and impacts across multiple scales—from immediate material effects to broader systemic changes—while acknowledging both opportunities and limitations in AI's environmental applications.

\subsection{``AI is a race to the bottom''}
A recent example comes from the film industry in the aftermath of the 2023 Hollywood union strikes where generative AI became a central dispute ~\cite{halperin2025filmworkers}. Since then, \textit{Lionsgate}'s partnership with AI startup \textit{Runway} has sent ripples through the professional community ~\citep{runway2024lionsgate_partnership}. The studio's decision to allow AI training on their film catalogue exemplified growing industry pressure to adopt AI for cost savings ~\citep{toonkel2024lionsgate_runway, landymore2024lionsgate_ai}. One can imagine how this market pressure affects individual studios: artists watch with growing concern as competitors begin using AI-generated content to underbid projects. When major clients threaten to leave, management hastily implements automated tools despite workers' explicit warnings about compromising professional craft and quality standards. Soon, the studio finds itself caught in a downward spiral—cutting rates to compete with AI-driven alternatives while experienced artists struggle to maintain quality standards against rushed, profit-driven implementations. This pattern, where market pressures drive rapid adoption of AI at the expense of quality and worker expertise, exemplifies how capitalist imperatives for profit maximization can override concerns about responsible development and professional craft. It drives a ``race to the bottom."

The designation of AI as a ``race to the bottom” has been a persistent critique in AI discourse, originating from concerns about recommender systems and social media platforms, where companies prioritized engagement at the expense of content quality and societal impact~\citep{harris_2017_b}. This critique finds parallels in other industries, such as food production, where rapid automation often led to the de-skilling of labor~\citep{robinson2016deskilling, robinson2019deskilling}, illustrating how technological adoption can disrupt professional expertise and established practices. Today, similar dynamics emerge in the AI sector, where major tech companie---Meta, OpenAI, Google, Anthropic, and others---release competing products in quick succession, often appearing to prioritize market presence over thorough testing or responsible deployment. This pattern amplifies broader anxieties about job displacement, safety risks, and the concentration of power in the hands of a few dominant corporations. As \citet{halperin2025filmworkers} argue, the Hollywood union workers recognized that filmmaking has always depending on evolving technology and thus did not so much oppose AI itself but rather its exploitative potentials as a cost-cutting measure.
This example underscores a critical insight: it is not the technology alone that drives negative outcomes, but the managerial structures and economic incentives shaping its deployment—factors that create a spiral of reduced costs and diminished standards, further fueling the ``race to the bottom.”

Yet, at the same time, this stopsign indeed stops productive conversations around how competitive pressure and resistance can help change the trajectory of AI development and implementation. The Hollywood union strikes in particular revealed how labor organizing and protest can ultimately lead to collective bargaining agreements that protect workers from the harms of AI~\citep{halperin2025filmworkers}. 
Meanwhile, in other filmmaking contexts, where not-for-profit incentives prevail, GenAI tools have shown potential to even lower barriers to entry and augment the creative capabilities of amateurs working outside the Hollywood system~\citep{halperin2025gen}. 
Similarly, open-source AI initiatives, operating under different organizational principles, often prioritize responsible development and democratized access over rapid commercialization~\citep{Costa2024}. While the use of AI at the end of the day still benefit the AI corporations, these varied contexts suggest that the so-called ``race to the bottom'' flattens how everyday users and workers have agency to chart alternative AI futures. The impact of market competition and labor organizing on AI development depends significantly on the underlying economic structures and governance frameworks---capable of driving more responsible or irresponsible practices depending on how power, incentives, and oversight are distributed.


\section{Proceeding with caution}
Overcoming semantic stopsigns requires a collective journey from surface-level reactions to deeper understanding and action. While the stopsigns often emerge from legitimate concerns such as environmental impact, bias, inequity, or reliability, they represent what \citet{scharmer2009theory} calls ``downloading'': falling back on habitual ways of thinking that prevent deeper inquiry. The challenge lies not in dismissing these concerns, but in using them as entry points for deeper individual and collective learning. This involves recognizing these cognitive traps to create space for new perspectives and nuanced approaches, whether they are for GenAI integration or resistance. By treating stopsigns as indicators of areas needing deeper exploration rather than endpoints, design communities can develop practices that acknowledge valid concerns while fostering meaningful innovation. In other words, we encourage AI critics to move thoughtfully beyond the stopsigns (rather than halt entirely) and AI evangelists to pause carefully (rather than race through them) such that, as a design research community, we collectively proceed with caution.

\subsection{Concrete strategies for critical engagement}
Design researchers and practitioners need concrete ways to move past semantic stopsigns toward more nuanced engagement with GenAI. The key is moving from abstract concerns to specific considerations that can inform day-to-day decision making. Our analysis reveals three complementary approaches to constructive engagement with AI challenges.

\subsubsection{Embracing productive tension: including diverse stakeholders}
Adopting a stance that balances skepticism with openness can allow design researchers and practitioners to benefit from \emph{healthy tension} and constructive conflict (e.g.~\citep{Kaila2024Agonistic}). For example, in one of our workshops, some attendees presented a case study on integrating an 'AI art' exhibit into a museum. To design the exhibit, they conducted a user study with museum visitors and digital technology specialists who worked there. This prompted one of us to inquire about the potential IP theft to the artists whose work it co-opts as training data and inquire why non-digital specialists and artists were not included in their design process. The authors said that it was because they suspected that such participants would be resistant to the use of AI. In this case, the stopsign attempted to 'stop' the harm to artists (see~\citep{jiang2023ai}). However, technology evangelists and users reportedly enjoyed the AI art exhibit. Thus, rather than halting the innovative project, a steady way past the stopsign could be to also recruit participants who are likely to push back against the use of AI (e.g., marginalized artists, art critics, and other AI skeptics). Deliberately seeking out anti-AI perspectives in AI design processes might lead to more ethical, balanced, and nuanced design decisions (e.g., \citet{halperin2023probing}).

\subsubsection{Reframing challenges as design opportunities}
Building on this inclusive approach, reframing apparent obstacles into design opportunities provides a complementary strategy. Rather than avoiding tension points—such as balancing efficiency gains against environmental impact or creative autonomy against AI assistance—they can be treated as design constraints that drive innovation. For instance, Shahryar Nashat's Reverse Rorschach transforms biometric data into evolving inkblot-like visuals, turning the unpredictability of GenAI into a dialogue between the artist, technology, and audience~\citep{reif2023reverseRorschach}. Similarly, the EU AI Act imposes restrictions such as prohibiting AI from detecting emotions~\citep{EUAIAct2024}. Investigating to what extent existing systems already do so may crystallize this debate, turning regulatory constraints into opportunities for clearer ethical guidelines~\citep{lomas2024improved}.
These examples highlight how ethical challenges and even so-called unreliable 'hallucinations' in generative systems can augment creativity~\citep{artificialdreams2024}, fostering thoughtful integration of GenAI that aligns with both innovative and ethical objectives.

\subsubsection{Documenting and sharing integration practices}
Another effective strategy focuses on systematic documentation of GenAI integration experiences, both successful and challenging. As Anthropic's Dario Amodei observes, when organizations document and share better practices for responsible AI development (such as Responsible Scaling Policies to align safety measures with AI capabilities~\citep{anthropic2023rsp} and mechanistic interpretability tools like Gemma Scope to enhance transparency~\citep{anthropic2023gemma}), it creates a 'race to the top' where innovations in responsible practice become competitive advantages that others can then adopt, ultimately raising standards across the field~\citep{fridman2024amodei}. An early example is the The World Economic Forum's Growth Summit 2023, which showcased numerous examples of how companies, like Heinz, have responsibly integrated GenAI into their practices~\citep{wef2023growth}. One such ripple effect is evident in the Global Lighthouse Network~\citep{wef2024lighthouse}, which embraced GenAI to transform sustainable manufacturing practices. This kind of impact is not limited to large-scale, institutional efforts; much of the progress in responsible AI adoption happens at the grassroots level, where individuals and small teams share best practices, iterate on processes, and inspire others to innovate responsibly. A recent study found that artists who documented and shared their GenAI practices helped create more democratized creative spaces~\citep{zhou2024generative}, while similar research demonstrated that teams who documented their GenAI writing processes helped others develop more nuanced approaches~\citep{doshi2024confronting}. Furthermore, it has been highlighted that transparent documentation of creative process with GenAI are necessary to gain research knowledge on the sustainability implications of such processes~\citep{jaaskelainen_environmental_2022}. 

\subsubsection{Developing reflective stances on GenAI} 
The last strategy addresses how design researchers and practitioners can move beyond semantic stopsigns by developing clear and context-sensitive perspectives on GenAI use. Reflecting systematically on key considerations provides the foundation for these perspectives, enabling practitioners to engage deeply with questions about GenAI's role in their work. This process not only enhances individual understanding but also facilitates meaningful conversations with colleagues, clients, and users. The framework of \textbf{reflective questions presented in Appendix \ref{sec:app}} exemplifies how practitioners can structure this process to yield actionable and well-articulated stances. It is a preliminary tool derived from workshop transcripts and group meeting discussions conducted during the development of this paper, requiring further validation and refinement in future studies. While the perspectives we outline here may not fit every scenario, they aim to provide clarity and guidance for integrating GenAI thoughtfully into design practice. Importantly, such stances are not intended to “wipe your slate clean” or erase critical tensions but rather to serve as a starting point for inquiry, building on the reflective questions you ask yourself. They provide a foundation for deeper engagement and iterative refinement, allowing practitioners to approach GenAI with clarity, purpose, and adaptability. To illustrate this approach, we, as the authors, propose the following stance as an example of how one might articulate a clear and reflective position on GenAI integration:

\begin{quotation}
\noindent We see GenAI as an opportunity to expand design knowledge while acknowledging its \textbf{material impact} on creative practice. Unlike traditional design tools, GenAI exhibits a relational influence---actively shaping creative direction through its ability to recombine existing knowledge and propose alternatives beyond immediate human intuition. This requires \textbf{developing new competencies} in dialogue with these systems, learning to critically evaluate their proposals and guide them toward meaningful outcomes. Embracing GenAI does not mean using it universally or wastefully, nor does selective non-use mean wholesale rejection. What matters is developing a considered relationship with these technologies, understanding when they serve design objectives and when they do not. Avoiding engagement altogether prevents exploration of deeper questions that emerge through practice---questions about the changing nature of design work and \textbf{human-AI collaboration}. Thoughtful experimentation and critical evaluation help build understanding needed for responsible integration. Environmental impact, IP, training data, and reliability are not universal blockers but considerations to evaluate within specific contexts---focusing on applications where impacts are justified by meaningful outcomes. The key is maintaining creative vision and critical judgment while developing competencies in working with these systems, understanding both their capabilities and limitations within a certain domain of practice.
\end{quotation}

\subsection{Limitations and future research directions}
This paper represents an initial attempt to advance conversations around GenAI in design research through our framework of ``semantic stopsigns." While aiming to move beyond oversimplified narratives, we recognize several key limitations of our approach. First, our findings are based on a limited set of workshops and interviews with self-selected participants who were likely already invested in or critical of GenAI. This focus may exclude perspectives from less-engaged domains or practitioners who operate under different constraints. Additionally, while we identified five stopsigns as key barriers in GenAI discourse, these represent only a subset of possible reductive framings. Other barriers or challenges likely remain unexplored, particularly in areas of design not represented in our data. The fast-evolving nature of GenAI presents another significant limitation. As new tools, ethical frameworks, and regulatory standards emerge, some of the challenges and opportunities discussed here may shift at a fast pace. Our cross-sectional approach, while providing valuable insights, captures only a specific moment in time and limits our ability to understand how long-term engagement with GenAI might reshape creative workflows, team dynamics, or professional identities. These limitations point to several critical research directions:

\begin{enumerate}
    \item A key opportunity lies in systematically analyzing how sustained GenAI use shapes creative practices over time, particularly how teams develop new workflows, evaluation criteria, and professional competencies. This calls for rich empirical studies examining how design organizations navigate the integration of these tools, complementing existing theoretical frameworks with grounded observations of practice.
    \item Another important area for exploration involves developing methods for evaluating the costs and benefits of GenAI in design processes. For instance, 
    being able to account for the short- and long-term sustainability impacts and their relation to each other (as discussed in Section \ref{sustainability}). In this regard, ~\citet{lupetti2025unbearablelightnesspromptingcritical} suggest translating sustainability impact into tangible metrics—such as “light bulb minutes” or “hamburgers eaten”---this can make the sustainability impacts of design decisions more accessible to non-expert audiences. The aim is not to detach or decontextualize these impacts but to provide relatable tools that can complement systemic and reflective approaches. Similar frameworks are needed for validating outputs and determining appropriate levels of human oversight across different design contexts~\citep[e.g.,][]{cavalcante2023meaningful}, ensuring both accessibility and responsible application. For example, when AI creates impossible building designs (Section \ref{sec:unrel}), designers need to review and correct them.
    \item Finally, there are opportunities to develop infrastructure for collective learning—shared case repositories, documentation standards, and practice-based communities. This knowledge-building cannot remain siloed in academic publications but must actively connect researchers with practitioners who are already developing ad-hoc solutions to these challenges. Their experiences should also inform both future research directions and practical guidelines for responsible GenAI integration.
\end{enumerate}

We view this work as an initial step toward addressing stagnation in GenAI discourse, aiming to spark more reflective and inclusive conversations rather than provide a definitive framework. Furthermore, this paper acts as a call for design researchers to engage with conflicting values, uncomfortable discussions, and varying perspectives -- because these are necessary to find knowledge and methods that can guide us forward in responsible, socially and environmentally just approaches to GenAI in design research.


\section{Conclusion}
This paper identifies five ``semantic stopsigns'' that can impede productive discourse and progress around GenAI in design research. By examining these reductive narratives and proposing concrete strategies for moving beyond them, we contribute a critical lens for the design research community to engage more deeply with the complex realities of GenAI integration in creative practice. Our approach of ``critical optimism'' offers a pathway forward, acknowledging valid concerns while still actively shaping the responsible development of AI in design. Thus, through this work we aim to empower design scholars and practitioners to surface and interrogate simplistic narratives in their own contexts, fostering more nuanced and reflective engagement with AI tools. As the field evolves, prioritizing open dialogue, experimentation, and knowledge-sharing will be crucial for realizing AI's potential while navigating its challenges. We call upon the design research community to confront these ``semantic stopsigns'' head-on, embracing the complexity and situatedness of GenAI's impact on creative work. By asking hard questions, sharing insights, and committing to ongoing reflection, we can collectively steer the integration of GenAI in design research towards more responsible and equitable futures.

\bibliographystyle{plainnat}  
\bibliography{00_references}  

\appendix
\section{Framework questions for GenAI in Design}
\label{sec:app}
\subsection{Understanding your context}
\begin{itemize}[label=\checkbox]
    \item What specific design tasks or processes are you considering using GenAI for?
    \item What are the core values and qualities that define your design practice/research?
    \item Who are the key stakeholders affected by your use of GenAI (users, clients, collaborators)?
\end{itemize}

\subsection{Technical understanding}
\begin{itemize}[label=\checkbox]
    \item Which specific models/tools are you considering using?
    \item Who developed these tools and what is known about their training?
    \item What are the documented capabilities and limitations of these tools?
    \item How will you validate and verify the outputs?
\end{itemize}

\subsection{Implementation considerations}
\begin{itemize}[label=\checkbox]
    \item How will GenAI integrate with your existing design processes?
    \item What level of human oversight and intervention is appropriate?
    \item How will you document and attribute AI contributions?
    \item What mechanisms will you use to evaluate success/failure?
\end{itemize}

\subsection{Resource and impact assessment}
\begin{itemize}[label=\checkbox]
    \item What computational resources are required for your intended use?
    \item How does this compare to alternative approaches?
    \item What are the environmental costs per use/project?
    \item Are there opportunities to optimize or reduce resource usage?
\end{itemize}

\subsection{Ethical considerations}
\begin{itemize}[label=\checkbox]
    \item How will you ensure transparency about AI use with stakeholders?
    \item What potential biases or limitations need to be monitored?
    \item How will you handle attribution and intellectual property considerations?
    \item What safeguards will you put in place?
\end{itemize}

\subsection{Evolution and adaptation}
\begin{itemize}[label=\checkbox]
    \item How will you stay informed about relevant developments?
    \item What triggers would cause you to reassess your approach?
    \item How will you document and share learnings?
    \item What mechanisms exist for updating your position?
\end{itemize}

\textbf{Note:} These questions are meant to be adapted and prioritized based on specific contexts. Not all questions will be relevant for every situation, and additional questions may be needed for particular cases.

\end{document}